\documentclass[aps,prl,twocolumn,reprint,superscriptaddress]{revtex4-1}  
\usepackage{graphicx}  
\usepackage{dcolumn}   
\usepackage{bm}        
\usepackage{amssymb}   
\usepackage{amsmath}
\usepackage{xcolor}
\definecolor{prlblue}{rgb}{0.176, 0.152, 0.57}
\usepackage[colorlinks=true, linkcolor=black, citecolor=prlblue, filecolor=black, urlcolor=prlblue]{hyperref}
\usepackage[mathlines]{lineno}

\begin{document}

\title{Emittance Preservation in an Aberration-Free Active Plasma Lens}

\author{C.~A.~Lindstr{\o}m}
\email{c.a.lindstrom@fys.uio.no}
\affiliation{Department of Physics, University of Oslo, 0316 Oslo, Norway}

\author{E.~Adli}
\affiliation{Department of Physics, University of Oslo, 0316 Oslo, Norway}
\author{G.~Boyle}
\affiliation{DESY, Notkestra{\ss}e 85, 22607 Hamburg, Germany}
\author{R.~Corsini}
\affiliation{CERN, CH-1211 Geneva 23, Switzerland}
\author{A.~E.~Dyson}
\affiliation{Department of Physics, University of Oxford, Clarendon Laboratory, Parks Road, Oxford OX1 3PU, United Kingdom}
\author{W.~Farabolini}
\affiliation{CERN, CH-1211 Geneva 23, Switzerland}
\author{S.~M.~Hooker}
\affiliation{Department of Physics, University of Oxford, Clarendon Laboratory, Parks Road, Oxford OX1 3PU, United Kingdom}
\affiliation{John Adams Institute for Accelerator Science, Denys Wilkinson Building, Keble Road, Oxford OX1 3RH, United Kingdom}
\author{M.~Meisel}
\author{J.~Osterhoff}
\author{J.-H. R{\"o}ckemann}
\author{L.~Schaper}
\affiliation{DESY, Notkestra{\ss}e 85, 22607 Hamburg, Germany}
\author{K.~N.~Sjobak}
\affiliation{Department of Physics, University of Oslo, 0316 Oslo, Norway}


\definecolor{light-gray}{gray}{0.75}
\renewcommand\linenumberfont{\normalfont\tiny\sffamily\color{light-gray}}

\begin{abstract}
Active plasma lensing is a compact technology for strong focusing of charged particle beams, which has gained considerable interest for use in novel accelerator schemes. While providing kT/m focusing gradients, active plasma lenses can have aberrations caused by a radially nonuniform plasma temperature profile, leading to degradation of the beam quality. We present the first direct measurement of this aberration, consistent with theory, and show that it can be fully suppressed by changing from a light gas species (helium) to a heavier gas species (argon). Based on this result, we demonstrate emittance preservation for an electron beam focused by an argon-filled active plasma lens.
\end{abstract}

\maketitle

Advances in high gradient acceleration research \cite{TajimaPRL1979,ChenPRL1985,RuthPA1985,JoshiPT2003} promise significantly more compact particle accelerators, key to next-generation free-electron lasers (FELs) \cite{CouprieJPB2014} and linear colliders \cite{LeemansPT2009}. However, advances in high gradient acceleration must be matched by a similar miniaturization of beam focusing devices. Active plasma lensing \cite{TilborgPRL2015} is one promising technique that provides compact, strong focusing in both planes simultaneously, by passing a large longitudinal current through a thin plasma-filled capillary \cite{SpencePRE2000,ButlerPRL2002}, ideally creating an azimuthal magnetic field proportional to the distance from the axis. While the concept dates back to the 1950s \cite{PanofskyRSI1950} and was used for fine focusing of heavy ion beams \cite{BoggaschAPL1992}, active plasma lenses (APLs) have recently gained attention based on their application to advanced accelerator research, such as beam capture and staging of laser plasma accelerators \cite{SteinkeNature2016}.

Although APLs provide kT/m focusing fields, orders of magnitude stronger focusing compared to conventional quadrupole magnets, they can suffer from aberrations that increase the emittance of the beam being focused \cite{TilborgPoP2018,PompiliAPL2017}. One such aberration is caused by plasma temperature gradients in the capillary (colder plasma closer to the wall), which leads to a radially nonlinear magnetic field distribution \cite{BobrovaPRE2001,BroksPRE2005} with enhanced focusing closer to the axis. This spherical aberration has been indirectly demonstrated in both helium \cite{TilborgPRAB2017} and hydrogen \cite{RoeckemannTBP2018}, by measurements of on axis field gradient enhancement and the formation of ring-shaped beams.

In this Letter, we show that this aberration can be fully suppressed by changing from a light gas species (helium) to a heavier gas species (argon). This discovery was made possible by the first complete characterization of the radial magnetic field distribution in an APL, in an experiment performed at the CERN Linear Electron Accelerator for Research (CLEAR) User Facility \cite{GambaNIMA2017,CorsiniIPAC2018}. The beam emittance was subsequently measured using quadrupole scans, resulting in the first demonstration of emittance preservation in an APL.

\begin{figure*}[t]
	\centering\includegraphics[width=0.98\textwidth]{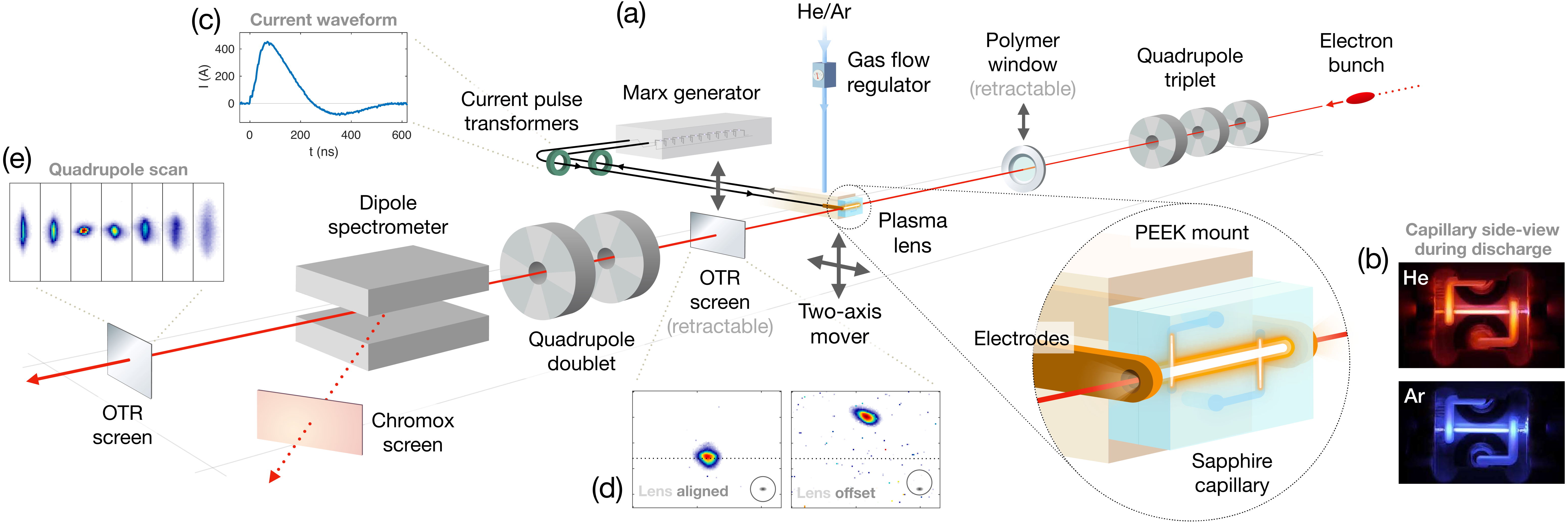}
	\caption{(a) Experimental setup: an electron bunch was tightly focused by a quadrupole triplet into an APL after passing a thin polymer window. The lens consisted of a gas-filled sapphire capillary with internal gas inlets (b) connected to an external gas flow regulator, and was discharged using two copper electrodes connected to a Marx generator producing high-voltage pulses with 410--450~A peak current (c). A two-axis mover scanned the beam transversely across the capillary aperture, deflecting the beam onto an OTR screen immediately downstream (d). With this screen retracted, the beam was instead focused by a quadrupole doublet onto another OTR screen, allowing measurement of emittance using a quadrupole scan (e). Additionally, a dipole spectrometer with a Chromox screen was used to measure the beam energy and energy spread.}
    \label{fig:ExptSetup}
\end{figure*}

The experimental setup \cite{LindstromNIMA2018}, shown in Fig.~\ref{fig:ExptSetup}, consisted of a 1~mm diameter and 15~mm long capillary milled from two sapphire blocks, mounted in the CLEAR beam line to allow passage of an electron beam. The capillary was filled with 1--100~mbar of gas through internal gas inlets, connected to an external flow regulator and a buffer volume. The gas escaping into the surrounding chamber was pumped out by a large turbo pump, which together with a 3~$\mu$m polymer (Mylar) window \cite{LishilinNIMA2016} preserved the ultrahigh vacuum in the upstream accelerator line. Holed copper electrodes on the up- and downstream side of the capillary were connected to a compact Marx bank \cite{DysonRevSci2016}, providing short 20~kV discharge pulses with a tunable 410--450~A peak current after 80~ns and a duration of 145~ns full width at half maximum (FWHM) [see Fig.~1(c)], as measured by in- and outgoing wideband current pulse transformers. A two-axis mover \cite{ToralEPAC2008} was used to displace the capillary horizontally and vertically relative to the beam, with a 1~$\mu$m step resolution and an approximate range of 10~mm.

To ensure a high-resolution magnetic field measurement, a quadrupole triplet 1~m upstream of the lens was used to focus the beam to a spot size of about $50\times50$~$\mu$m root mean square (rms). This was measured and optimized at the plasma lens using optical transition radiation (OTR) from a stainless steel wedge mounted on the upstream electrode. Directly downstream of the lens (30 cm) was a retractable OTR screen to observe beam focusing and centroid angular deflections from the APL, mounted with a thin aluminum foil to block stray plasma light. Further downstream, a quadrupole doublet allowed multishot emittance measurements using quadrupole scans on another OTR screen, also with a noninvasive light-blocking foil. A dipole magnet was used as a spectrometer to measure the mean energy (200--220~MeV) and energy spread ($<$0.2\% rms) of the beam on a chromium-doped ceramic (Chromox) screen. Upstream of the experimental setup was a radio frequency transverse deflecting cavity used to measure the bunch length to be approximately 400~$\mu$m rms, as well as toroids used to measure the beam charge.

The measurement of the radial magnetic field distribution in the APL was performed by displacing the lens vertically across the full 1~mm aperture of the capillary with respect to a tightly focused fixed-orbit beam, while centered in the horizontal plane. Angular deflections of the beam centroid, as observed on the closest OTR screen, scale linearly with the local magnetic field averaged over the length of the capillary. A short capillary was therefore used to avoid any transverse displacement (betatron motion) inside the APL, as this would lead to unwanted radial averaging. Each offset was recorded over 5--10~shots to average any angular jitter, which was approximately 0.5 and 0.1~mrad rms in the horizontal and vertical plane, respectively. The scans were performed around the peak current timing (after approximately 80~ns), as this is the most stable and potent operating point and because later timings with lower discharge current tended to suffer from poor signal-to-noise ratio. One or two bunches (at a 667~ps interval) with 5--7~pC of charge per bunch were used to simultaneously ensure negligible distortion from plasma wakefields \cite{LindstromPRAB2018} and to get a sufficient signal on the OTR screen.

The expected magnetic field in an APL can be found using Amp{\`e}re's law for a longitudinal current density,
\begin{equation}
	\label{eq:AmperesLaw}
	\frac{1}{r} \frac{\partial}{\partial r} \left( r B_{\phi} \right) = \mu_0 J_z(r),
\end{equation}
where $B$ is the magnetic field, $J$ is the current density, the permeability of the plasma is assumed to be that of the vacuum $\mu_0$, and $r$, $\phi$, and $z$ are the radial, azimuthal, and longitudinal coordinates, respectively. If the current density is uniform, Eq.~(\ref{eq:AmperesLaw}) integrates to give a linear magnetic field with a constant magnetic field gradient
\begin{equation}
	\label{eq:GradientUniformCurrent}
	g_r = \frac{\partial B_{\phi}}{\partial r} = \frac{\mu_0 I_0}{2 \pi R^2},
\end{equation}
where $I_0$ is the total current and $R$ is the capillary radius. This represents the ideal operation of an APL, providing emittance preservation and focusing in both planes.

However, this picture is complicated by the buildup of a radial temperature gradient inside the capillary, which leads to a nonuniform current density and a nonlinear magnetic field---detrimental to the beam quality. As described in Ref.~\cite{BroksPRE2005} and supported by Ref.~\cite{GonsalvesPRL2007}, this occurs in a four-step process, starting with (1) the formation of a cold plasma. Then (2) the electron temperature increases sharply from Joule heating, but (3) due to a thin, virtually electron-free sheath near the capillary wall, the hot electrons only transfer their heat to the plasma ions, which (4) subsequently lose heat to the wall. This process preferentially cools the plasma closer to the capillary wall, leading to the formation of a nonuniform temperature profile with hotter plasma closer to the axis. Since the plasma conductivity $\sigma$ increases with the plasma electron temperature $T_e$, the current concentrates closer to the axis, as given by \cite{TilborgPRAB2017}
\begin{equation}
	\label{eq:CurrDensTempScaling}
	J_z(r) = \sigma(r) E_z \propto T_e^{3/2}(r),
\end{equation}
where $E_z$ is a uniform longitudinal electric field. 

A steady-state solution to this process was found by Ref.~\cite{BobrovaPRE2001} through a simplified magnetohydrodynamics (MHD) approach, satisfying the radial heat flow equation
\begin{equation}
	\frac{1}{x} \frac{\partial}{\partial x} \left( x \frac{\partial u}{\partial x} \right) + u^{3/7} = 0,
\end{equation}
where $x = r/R$ is a scaled radius and $u = (T_e/A)^{7/2}$ is a scaled temperature for which $A = \sqrt{7 R^2 E_z^2 \sigma_0 / 2 \kappa_0}$. Here we assume a Maxwellian velocity distribution, such that the thermal and electrical conductivities scale according to $\kappa = \kappa_0 T_e^{5/2}$ and $\sigma = \sigma_0 T_e^{3/2}$, respectively \cite{SpitzerPR1953}. Substituted into Eq.~(\ref{eq:CurrDensTempScaling}), we find the current density profile
\begin{equation}
	J_z(r) = \frac{I_0}{\pi R^2} \frac{u(r)^{3/7}}{2 m_I},
\end{equation}
where the scaled temperature is normalized by
\begin{equation}
	m_I = \int_0^1 u(x)^{3/7} x dx,
\end{equation}
to ensure the correct total current $2 \pi \int_0^R J_z(r) r dr = I_0$. With this current density, Amp{\`e}re's law (\ref{eq:AmperesLaw}) can be numerically integrated to find the steady-state radial magnetic field distribution---sometimes termed the ``JT model.'' A nonuniformity will lead to an enhancement of the on axis focusing gradient 1--1.48 times larger than Eq.~(\ref{eq:GradientUniformCurrent}), depending on the wall temperature.

In order to avoid the nonuniformity, we must break the assumption of steady state. In a light gas, this is not trivial, as the timescale of electron--ion heat transfer, and hence the buildup of the nonuniformity, is shorter than the typical rise time of the current pulse. However, crucially, this timescale can be slowed down by changing to a heavier gas, where the rate of thermal transfer between electrons and ions as well as the ion thermal conductivity (both inversely proportional to the ion mass \cite{BobrovaPRE2001}) are significantly reduced. The discharge current can then rise to its peak before the current becomes nonuniform, ensuring a linear magnetic field when the beam passes. Two-temperature MHD simulations using \textsc{flash} \cite{Fryxell2018} are currently under study to verify this explanation and will be the subject of a future publication.

\begin{figure}[b]
	\centering\includegraphics[width=0.93\linewidth]{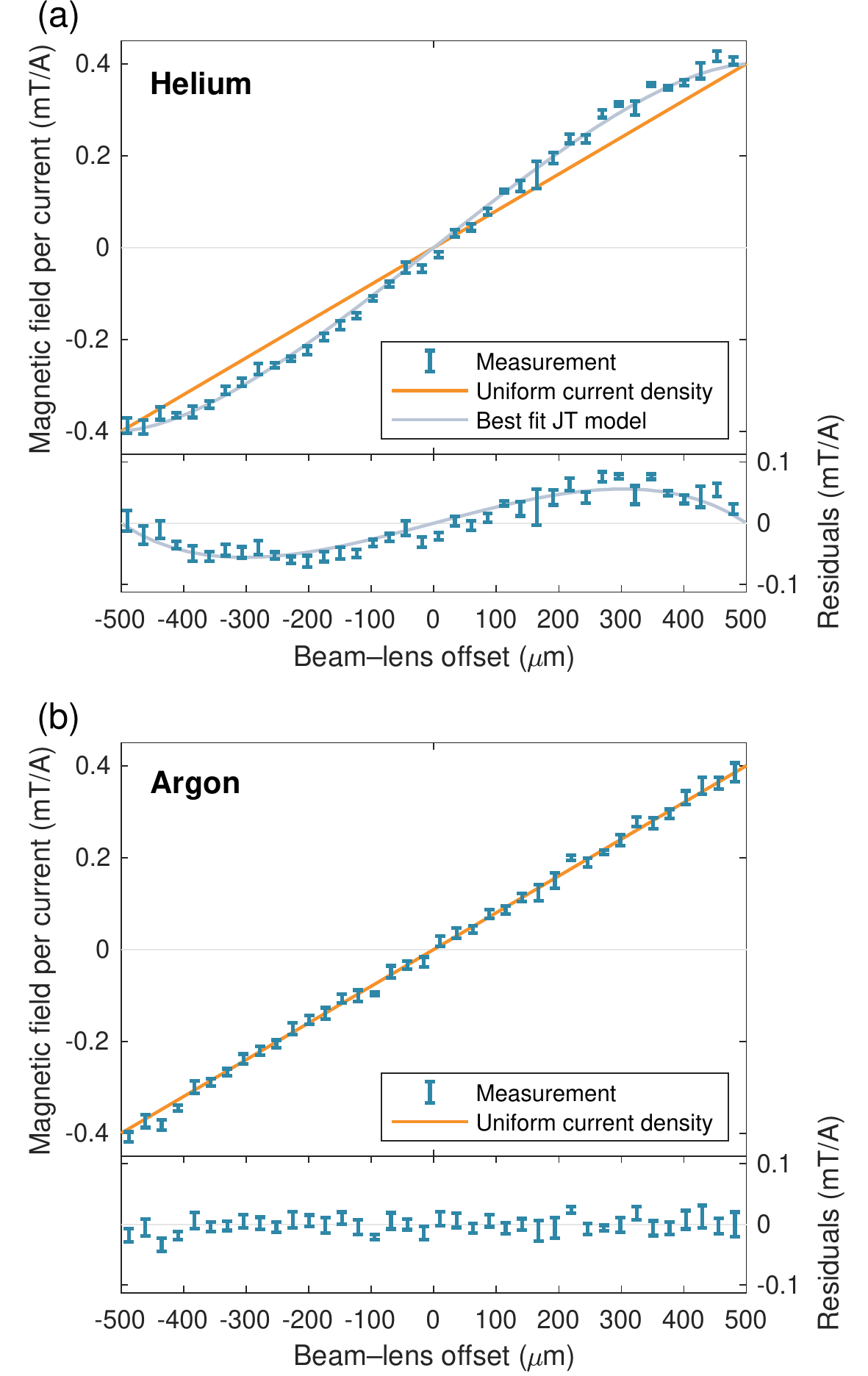}
	\caption{Measurement of the magnetic field per discharge current for a scan of beam-to-lens offsets in (a) helium and (b) argon, where the uncertainty (blue error bars) represents the standard deviation of the mean. A strong nonlinearity is observed in helium, consistent with the JT model (gray line), whereas in argon the measurement is consistent with the expectation from a uniform current density (orange lines).}
    \label{fig:OffsetScans}
\end{figure}

Experimentally, this magnetic field distribution was found by measuring the angular deflection of the beam as an offset $\Delta y_{\mathrm{OTR}}$ on the downstream OTR screen for every offset $y_0$ of the lens. Since the current in the APL was fluctuating by a few percent, the measurement can be improved by considering the ratio of the magnetic field and the instantaneous discharge current observed by the beam
\begin{equation}
	\label{eq:BfieldPerCurrent}
	\frac{B_{\phi}(y_0)}{I_0} = \frac{E \Delta y_{\mathrm{OTR}}}{e c L \Delta s I_0},
\end{equation}
where $L$ is the length of the capillary, $\Delta s$ is the distance from the center of the capillary to the screen, $E$ is the beam energy, and $e$ and $c$ are the electron charge and the speed of light in vacuum, respectively.

\begin{figure*}[t]
	\centering\includegraphics[width=\linewidth]{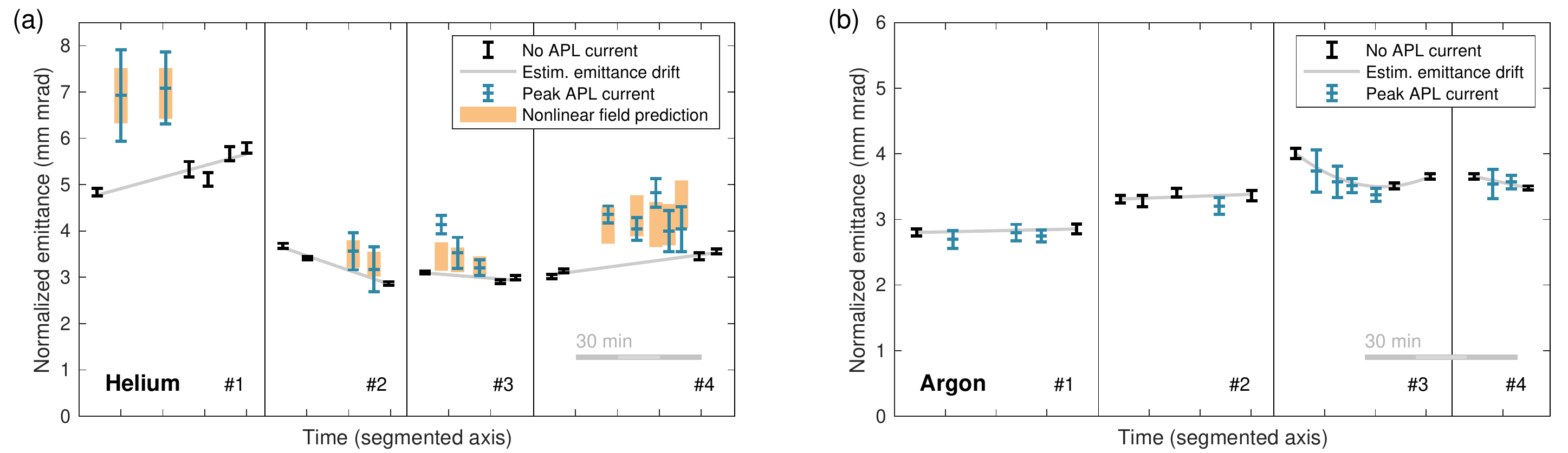}
	\caption{Quadrupole scan emittance measurements for both (a) a helium and (b) an argon APL, performed multiple times at peak current timing (blue error bars). Additionally, the background emittance was measured in the absence of current (black error bars) before and after discharges to estimate any emittance drift (gray lines). The predicted emittance growth in helium (orange rectangles) based on the measured nonlinear field [see Fig.~\ref{fig:OffsetScans}(a)] is in good agreement with the measured values. In argon, all measurements are consistent with emittance preservation. Emittance drift is modeled with a linear fit in all measurements except one (argon \#3), where a quadratic fit produces a tighter bound.}
    \label{fig:EmittanceScans}
\end{figure*}

Figure~\ref{fig:OffsetScans} shows the measured magnetic field per current for both (a) helium and (b) argon, using a transverse step size of 26~$\mu$m. In helium, there is clear evidence of a nonlinearity, consistent with the JT model and indicating a best fit gradient enhancement factor of 1.34 and a scaled wall temperature $u(R)$ = 0.0114. These results are in excellent agreement with Ref.~\cite{TilborgPRAB2017}. In argon, there is no evidence of any nonlinearity---the magnetic field distribution is linear to within the error of the measurement. The flow of each gas was minimized while ensuring stable discharges at the few nanosecond level. The resulting neutral gas density in the capillary was 6~mbar in argon and 23~mbar in helium, both a 70\% pressure drop from the buffer volume, measured by sealing one gas inlet and connecting the closest end of the capillary to a capacitance gauge---a method used also in Ref.~\cite{TilborgOL2018}.

To verify the expected emittance growth in helium and emittance preservation in argon, a number of quadrupole scans were performed in each gas. Instead of using a tightly focused beam, a larger beam (100--150~$\mu$m rms) covering a significant portion of the aperture was used---this way, the nonlinearity was sampled more extensively and the potential emittance growth increased. Simultaneously, to avoid any emittance growth from plasma wakefields, the beam charge was lowered to approximately 2~pC for the single bunch used in the measurement. Due to non-negligible horizontal dispersion, emittance measurements were only performed in the vertical plane. Additionally, for each measurement, at least two different current settings were used in the second (nonscanned) quadrupole, allowing an overall verification of length and current calibrations.

Figure~\ref{fig:EmittanceScans} shows emittance measurements from multiple quadrupole scans in both (a) helium and (b) argon, repeated four times for each gas. Each segment consists of one or more control measurements before and after the shots with discharge to estimate any emittance drift over a 15--30~min interval, as well as several emittance measurements where the beam is focused by the APL at peak current (410~A). We clearly observe emittance growth in helium compared to the background emittance, in good agreement with predictions from particle tracking through the measured nonlinear field [see Fig.~\ref{fig:OffsetScans}(a)]. This tracking simulation uses the measured spot size in the lens as well as a random centroid offset jitter (estimated to $1\sigma$ beam size), leading to a spread of predicted emittances as more offset beams sample the nonlinearity more strongly. The emittance error in each quadrupole scan is obtained from the covariance matrix produced when performing parabolic fits to the measured spot sizes. This error is observed to increase during discharges, both due to the centroid offset jitter as well as current fluctuations caused by a discharge timing jitter.

In argon, the measured emittance during peak discharge current is fully consistent with the background emittance to within the estimated error. This is clear evidence of emittance preservation, simultaneously confirming that there are no other sources of emittance growth. Assuming that additional emittance is added in quadrature and that errors are Gaussian, the argon measurement excludes emittance growth larger than 0.25~mm~mrad at 90\% confidence level. Moreover, the change in beam optics was measured across consecutive on-off quadrupole scans to be consistent with the expected focusing from a 326~T/m uniform magnetic field gradient [Eq.~(\ref{eq:GradientUniformCurrent})] to within the error of the measurement, verifying that there is no gradient enhancement in argon.

We have shown that APLs can be made aberration-free by changing to a heavier gas species, but this comes at the cost of more scattering. Emittance growth from multiple Coulomb scattering \cite{ZimmermannJPCS2017,KirbyPAC2007} increases almost quadratically with atomic number, such that argon scatters 54 times more than helium and 280 times more than hydrogen. This effect can, however, be minimized by increasing the discharge current or decreasing the capillary radius, thereby requiring a shorter lens for the same focusing or by lowering the pressure. For this experiment, the pressure was sufficiently low to not increase the emittance, as verified by quadrupole scans with and without gas, but calculations indicate that higher pressures could result in non-negligible emittance growth. Moreover, scattering can potentially be reduced by using an intermediate gas species, like nitrogen or neon, if the aberration can still be suppressed. Use of nitrogen, which scatters 5.6 times less than argon, is currently a topic of active investigation.

One immediate application of the argon lens is as an emittance preserving beam capture device for laser plasma accelerators (LPAs). A challenge for LPAs is the highly diverging beams produced, typically 1~mrad rms, which combined with percent-level energy spreads lead to significant emittance growth due to large chromaticity during beam capture. This problem can be solved by using an aberration-free active plasma lens (e.g., 600~A peak current, 10 mm long, 400~$\mu$m capillary radius, 1~mbar argon) placed sufficiently close to a LPA source (10~cm downstream) to capture high-quality beams without degradation (1~mm~mrad, 200~MeV, 1\% rms energy spread, 1--2~$\mu$m rms bunch length, up to 200~pC)---potentially useful for an ultracompact FEL.

Other applications may include radially symmetric final focusing for linear colliders or possibly staging of plasma accelerators \cite{LindstromNIMA2016}, assuming plasma wakefield distortion is avoided by reducing the beam intensity. While in this measurement plasma wakefields were successfully controlled for, in general, they will impose limits to the application of APLs to low-emittance, high-intensity beams \cite{LindstromPRAB2018}, such as those needed for linear colliders, unless compensation methods can be found. The nonlinearity suppression reported in this Letter contributes in this regard by increasing the effective aperture of the lens, allowing significant reduction of wakefields with the use of larger, lower density beams.

In conclusion, emittance preservation in an active plasma lens has been demonstrated for the first time with the use of an argon-based discharge capillary. Direct measurements of magnetic fields across the full aperture show linearity in argon and nonlinearity in helium. Quadrupole scans demonstrate the expected emittance preservation and growth, respectively, consistent with the measured field profiles. This development of a compact linear beam optics device is a critical step towards truly compact, low-emittance accelerators.

\begin{acknowledgments}
The authors wish to thank Davide Gamba, Alessandro Curcio, Reidar Lunde Lillest{\o}l, Gianfranco Ravida, Gerard McMonagle, Franck Perret, Stephane Curt, Thibaut Lefevre, Bruno Cassany, Serge Lebet, Jose Antonio Ferreira Somoza, Alice Ingrid Michet, Herv\'e Rambeau, Stefano Mazzoni, Michael John Barnes, Aimee Ross, and Candy Capelli. We thank CERN for providing beam time at the CLEAR User Facility. This work was supported by the Research Council of Norway (NFR Grant No.~230450) and by the Helmholtz Association of German Research Centers (Grants No.~VH-VI-503 and No.~ZT-0009).
\end{acknowledgments}


\end{document}